\documentclass{agujournal}
\usepackage{natbib}
\usepackage[shortlabels]{enumitem}

\journalname{JGR-Space Physics}

\begin{document}
	\title{Electron acceleration and thermalization at magnetotail separatrices}	
	
	\authors{C. Norgren\affil{1},
		M. Hesse\affil{1},
		P. Tenfjord\affil{1},
		D. B. Graham\affil{2},
		Yu. V. Khotyaintsev\affil{2},
		A. Vaivads\affil{3},
		K. Steinvall\affil{2,4},
		Y. Xu \affil{5},
		D. J. Gershman\affil{6,7},		
		P.-A. Lindqvist\affil{3},
		J. L. Burch\affil{8}
	}
	\affiliation{1}{Birkeland Centre For Space Science, University of Bergen, Bergen 5007, Norway}
	\affiliation{2}{Swedish Institute of Space Physics, Uppsala 75121, Sweden}			
	\affiliation{3}{Space and Plasma Physics, School of Electrical Engineering, KTH Royal Institute of Technology, Stockholm 11428, Sweden}
	\affiliation{4}{Space and Plasma Physics, Department of Physics and Astronomy, Uppsala University, Uppsala 751 20, Sweden}
	\affiliation{5}{School of Space and Environment, Beihang University, Beijing, China}
	\affiliation{6}{NASA Goddard Space Flight Center, Greenbelt, Maryland 20771, USA}
	\affiliation{7}{Department of Astronomy, University of Maryland, College Park, Maryland 20742, USA}
	\affiliation{8}{Southwest Research Institute, San Antonio, Texas 78228, USA}
	
	\correspondingauthor{Cecilia Norgren}{cecilia.norgren@uib.no}

	\begin{abstract}		
		In this study we use the Magnetospheric Multiscale (MMS) mission to investigate the electron acceleration and thermalization occurring along the magnetic reconnection separatrices in the magnetotail. We find that initially cold electron lobe populations are accelerated towards the X line forming beams with energies up to a few keV's, corresponding to a substantial fraction of the electron thermal energy inside the exhaust. The accelerated electron populations are unstable to the formation of electrostatic waves which develop into nonlinear electrostatic solitary waves. The waves' amplitudes are large enough to interact efficiently with a large part of the electron population, including the electron beam. The wave-particle interaction gradually thermalizes the beam, transforming directed drift energy to thermal energy. 
	\end{abstract}

	\section{Introduction}	
	Magnetic reconnection is a universal process where magnetic energy is often explosively released, leading to particle acceleration and heating. Observations suggest that magnetic reconnection and subsequent processes can accelerate electrons to energies of tens or even hundreds of keV’s \citep[e.g.][]{Hoshino2001,Oieroset2002,Vaivads2011,Fu2019}. The particle energization associated with magnetic reconnection is known to take place in several regions: in the inflow region and along the separatrices \citep[e.g.][]{Nagai2001,Egedal2008,Hesse2018_separatrix,Eriksson2018}, inside the ion and electron diffusion regions \citep[e.g.][]{Torbert2018,Hesse2018_rec_efield,Wang2018_energyconversion}, in the magnetic reconnection exhaust \citep[e.g.][]{Bessho2015}, in the vicinity of magnetic islands \citep{Chen2008a,Huang2012}, both during island coalescence \citep{Pritchett2008_pop} and contraction \citep{Drake2006}, and at dipolarization fronts \citep[e.g.][]{Fu2011,Vaivads2011}. Where, how, and to what extent the particles are accelerated depend not only on fundamental properties such as the particle species and the relative composition of species, but also on changing properties, such as the particle's velocity. Two examples of the former is that the presence of heavier ions or cold ionospheric ions can act as energy sinks in addition to reducing the rate at which magnetic flux is being reconnected \citep[e.g.][]{ToledoRedondo2017_energybudget,Tenfjord2019}. One example of the latter is Fermi type B acceleration in the reconnection exhaust or in magnetic islands where the energization is more efficient if the initial velocity is higher \citep{Northrop1963,Drake2006}. A clear indication of the non-uniform energization of a particle species is the fact that the energy partition is generally not uniform, with some particles being accelerated to superthermal energies, while some remain thermal \citep[e.g.][]{Hoshino2001}. How this energy-dependent energization affects the bulk energization of a species is unclear. For example, a study of the change in electron temperature between the magnetosheath and the reconnection exhaust during reconnection at the dayside magnetopause did not show any strong dependence on the initial electron temperature in the magnetosheath \citep{Phan2013}. The acceleration mechanisms can vary between direct acceleration by electric fields, for example the reconnection electric field inside the diffusion regions \citep{Bessho2015}, the already mentioned Fermi acceleration \citep{Fermi1949}, and betatron acceleration \citep[e.g.][]{Northrop1963}. Ultimately, due to the conservation of energy, the final plasma energies must depend on the amount of available magnetic energy compared to the amount of plasma to be reconnected, which varies during the reconnection process \citep{Vaivads2011,Ergun2018}.

	In addition, energy transfer does not always occur directly between the magnetic field and the particles, but often in steps, between different plasma populations, mediated by electromagnetic fields. One such example is the Hall magnetic fields, which are due to the different motions of ion and electrons. Observations from the terrestrial magnetotail show that at the separatrices of magnetic reconnection, lower-energy field aligned electrons flow into the reconnection region while higher-energy electrons flow out of the reconnection region \citep{Oieroset2001,Nagai2001,Asano2008} carrying the outward and inward Hall currents, respectively \citep{Nagai2003}. The acceleration leading to the formation of these electron flows and associated currents has by some authors been suggested to be a necessity to maintain quasi-neutrality inside the ion diffusion region \citep[][]{Uzdensky2006,Egedal2008}. It has been explained as following: inside the ion diffusion region, the demagnetized ions are free to move across the magnetic field while the magnetized electrons are tightly bound to the magnetic field lines. As a magnetic flux tube expands while convecting inwards, the ion density can thus remain close to constant while the electron density must decrease. The resulting charge separation produces an electric field that accelerates electrons inward, which can lead to the formation of beams and temperature anisotropies \citep{Egedal2005}. In some cases, the electric field can become localized leading to the formation of double layers \citep{Ergun2009,Wang2014,Egedal2015}. The ultimate effect of electron acceleration along the separatrices remains disputed. For example, \cite{Bessho2015} found that the separatrix acceleration occurring at the inbound leg of an electron trajectory was mostly negated by the decelerating effect when the same electron arrived close to the separatrix on the opposite side of the neutral sheet. Meanwhile, \cite{Egedal2015} argued that the confining nature of the potential could lead to more efficient energization within the exhaust by the reconnection electric field. Furthermore, as mentioned above, an initially higher electron velocity would also lead to more efficient Fermi acceleration.
		
	On a more local scale, electromagnetic waves can mediate energy transfer between different plasma populations. For example, the counter-streaming hot and cold electron populations occurring at reconnection separatrices has been studied extensively with numerical simulations. They have been shown to be unstable to the generation of electrostatic waves, leading to the thermalization of the cold electron beam \citep{Divin2012,Fujimoto2014,Huang2014,Egedal2015,Chen2015_sepwaves}. Depending on the velocity at which the waves are generated, they can interact with different parts of the electron distributions \citep{Omura1996,Graham2015a}, and transfer energy between them. In the nonlinear stages of instabilities, it is common that electron trapping by the strong wave potential leads to electron phase space holes and electrostatic solitary waves (ESWs) \citep[e.g.][]{Mozer2018}. At reconnection separatrices, the interface between the inflowing and outflowing electrons represents a velocity shear. In such an environment, the instabilities developing may lead to transfer of energy not only between different energy ranges, but also across the boundary \citep{Hesse2018_separatrix}. Although electrostatic waves and ESWs are commonly observed at reconnection separatrices in conjunction with electron beams \citep{Cattell2005,Viberg2013} or plateau distributions associated with significant drift speed \citep{Ergun1998b}, their effect on plasma populations has not been firmly established.

	In order to determine the importance of the separatrix acceleration and subsequent wave-particle interaction for the overall electron energization during magnetic reconnection, it is necessary to observe these phenomena in space. %Historically, the necessary measurements to investigate this have not been possible. 
	In this study we will do so, using high-cadence plasma and field measurements by the four closely separated Magnetospheric Multiscale Mission (MMS) spacecraft. We are able to make detailed measurements of both the electron acceleration and subsequent wave-particle interaction at separatrix regions in the magnetotail. 
	
	\section{Observations}	
	In this section, we report MMS observations from the plasma sheet boundary layer. The electric field is from the Electric field Double Probes (EDP) \citep{Lindqvist2014,Ergun2014}, the magnetic field is from the FluxGate Magnetometer (FGM) \citep{Russell2014}, the plasma distributions and moments are from the Fast Plasma Investigation (FPI) \citep{Pollock2016}. All times are given in Coordinated Universal Time (UTC). Unless otherwise stated, positions and vectors are given in Geocentric Solar Ecliptic (GSE) coordinates.
	
	We will first present observations of relatively thin channels of electron jets directed opposite to the broader ion and electron flows that is the exhaust flow of magnetic reconnection. For a few events, we will quantify the level of acceleration and compare it to the thermal energy of the lobe population and the plasma sheet population. We will then investigate the wave activity within these regions of accelerated electrons to infer the wave-particle interaction. We focus on electrostatic waves that propagate predominantly along the ambient magnetic field.
			
	\section{Electron acceleration channel}
\begin{figure}	
		\includegraphics[width=1\textwidth]{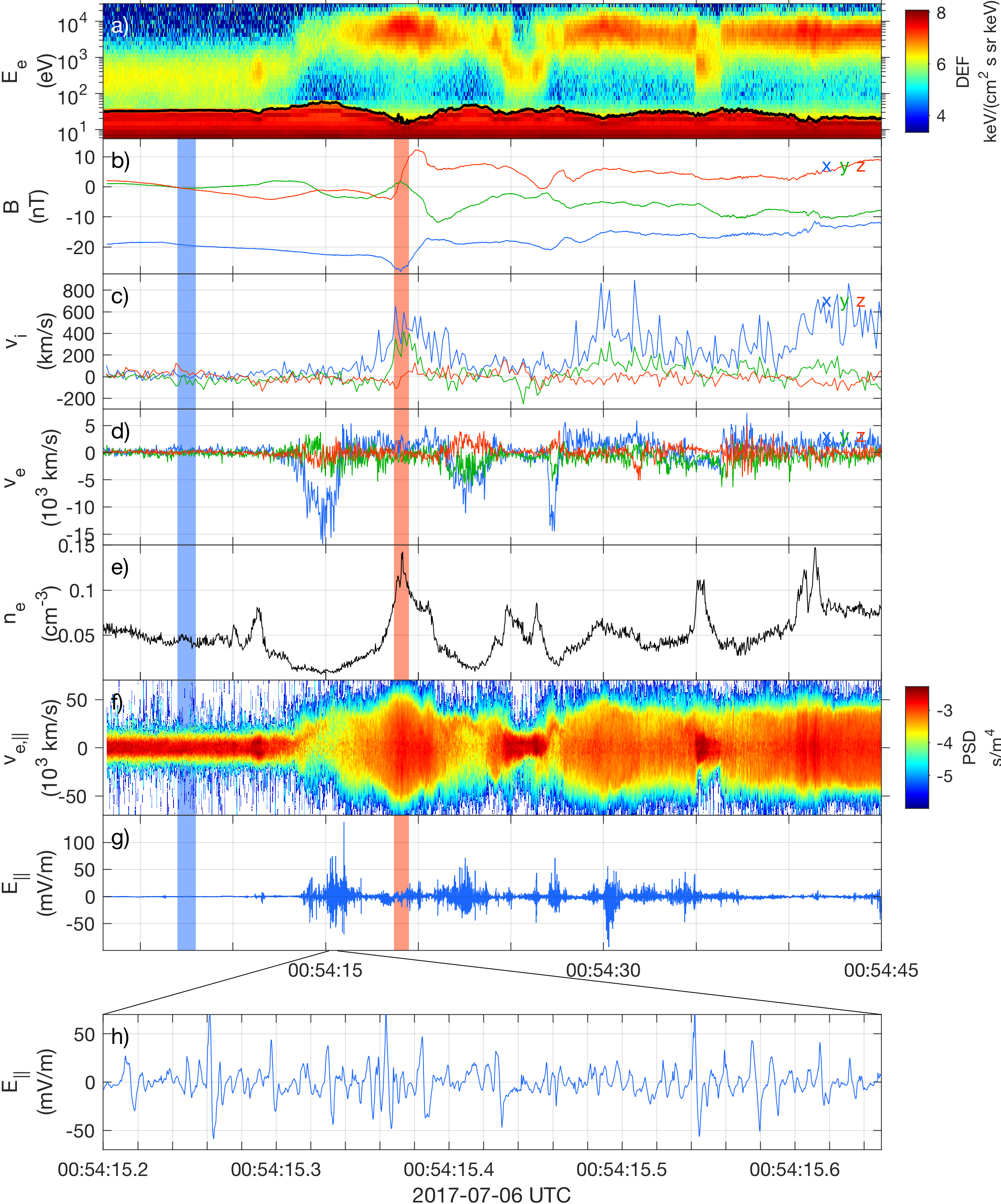}
		\caption{Overview of separatrix crossing. The blue and red shaded areas indicate the time intervals from which we extract lobe and plasma sheet parameters, respectively. (a) Energy spectrogram of electron differential energy flux. The black line shows the spacecraft potential, below which spacecraft photoelectrons are present. (b) Magnetic field. (c) Ion velocity. (d) Electron velocity. (e) Electron density. (f) Reduced electron phase space density distribution, integrated over the directions perpendicular to the magnetic field. (g)-(h) Parallel electric field.}
		\label{fig:separatrix_environment}
	\end{figure}

	In this section we present an event from the magnetotail observed on June 7, 2018 at $[-18,4,2]$ Earth radii, investigated previously by \cite{Huang2019}. Figure \ref{fig:separatrix_environment}a shows the electron differential energy flux (DEF), in which we can identify the lobe at lower energies ($E_e\lesssim1$ keV), and the plasma sheet at higher energies ($E_e\gtrsim1$ keV). The spacecraft are initially located in the southern lobe before they enter the plasma sheet boundary layer and the outer edges of the plasma sheet. The spacecraft make partial exits into the lobe two more times before residing in the plasma sheet until the end of the displayed time interval. During this time, the magnetic field is predominantly tailward ($B_x<0$). However, at 00:54:20 when the spacecraft encounters the plasma sheet for the first time, a significant northward component ($B_z>0$) appears, closely associated with changes in $B_x$ and $B_y$. \cite{Huang2019} interpreted this as a passing flux rope. The ion flow is Earthward ($v_{ix}>0$), indicating the spacecraft are located in the Earthward exhaust of a magnetic reconnection X line. We note that although $v_{ix}$ maximizes at $\sim 800$ km/s during the shown interval, the ion distribution consists of two populations: one cold population with bulk speed close to $v_{ix}=0$ km/s, and another hotter population streaming Earthward at speeds $>1000$ km/s (not shown). At later times (not shown), the ion flow reverses, indicating that the X line is moving tailward. The Earthward ion flow is matched by an Earthward electron flow ($v_{ex}>0$). However, at the edges of the Earthward flow, three shorter intervals of larger amplitude tailward ($v_{ex}<0$) flows are observed. The electron flows are consistent with the current derived from the magnetic field (not shown). These regions are associated with density cavities where $n_e\sim0.01$ cm$^{-3}$ (Figure \ref{fig:separatrix_environment}e). In comparison, the density in the lobes (before 00:54:10) is $n_{e}^{lb}\sim 0.05$  cm$^{-3}$, and the largest density during the interval, which we associate with a plasma sheet population, is $n_e^{sh}\sim0.15$ cm$^{-3}$. We shall henceforth refer to the regions of low densities and large amplitude electron flows as acceleration channels. 
	
	Figure \ref{fig:separatrix_environment}f shows the phase space density (PSD) of the reduced electron distribution projected onto the magnetic field:
	\begin{eqnarray*}
	f_{e}^{1D}(v_{\parallel}) = \int_{-\infty} ^{-\infty} f_{e}(v_{\parallel},v_{\perp,1},v_{\perp,2})dv_{\perp,1}dv_{\perp,2}.
	\end{eqnarray*}
	Inside the three acceleration channels, we can clearly see the accelerated population for $v_{e\parallel}>0$. These acceleration channels with electron flows directed towards the magnetic reconnection X line, opposite to the exhaust flow, are prominent features seen in the separatrix regions in numerical simulations of magnetic reconnection \cite[e.g.][]{Divin2012,Fujimoto2014,Egedal2015,Hesse2018_separatrix}. The reduced distributions presented in the simulations bear strong resemblance to the reduced electron distributions observed by MMS presented here. 
	
	All the acceleration channels occur at the edges of the reconnection outflow, where we expect the separatrices to be. Therefore, consistent with numerical simulations, we identify the acceleration channels as being located at the separatrices of a magnetic reconnection site. However, from numerical simulations, we know that density cavities associated with the accelerated populations do not always extend over the whole length of the separatrices \citep[e.g.][]{Egedal2015}. In addition, in the presence of a guide field the acceleration regions are to some degree suppressed at two opposing of the four separatrices \citep[e.g.][]{Pritchett2004}. Therefore, while accelerations channels and reconnection separatrices are closely related, they are not always coincident. Also, importantly, electron acceleration channels are not exclusively related to magnetic reconnection but can occur in a multitude of plasma environments.
	
	 Since we do not observe any reconnection outflow reversals in the immediate proximity of the acceleration channels, we have no straightforward means of determining how far away from the X line they are located. 		 
	
	The acceleration channels are associated with large amplitude parallel (Figure \ref{fig:separatrix_environment}g-\ref{fig:separatrix_environment}h) and perpendicular (shown for a shorter time interval in Figure \ref{fig:acc_channel_thickness}) electric field fluctuations. Since instabilities driven by parallel beams often result in large amplitude parallel electric fields, we will in sections \ref{sec:waves} and \ref{sec:instability} focus on investigating the relation between the field-aligned electric fields and the parallel streaming populations. First, however, we will quantify in more detail the electron acceleration.
	
	\section{Acceleration potential}	
	\label{sec:acceleration_potential}
\begin{figure} % +table.separatrix_acceleration
% sep_acc.figure2_get_acc_pot.m	
		\includegraphics[width=1\textwidth]{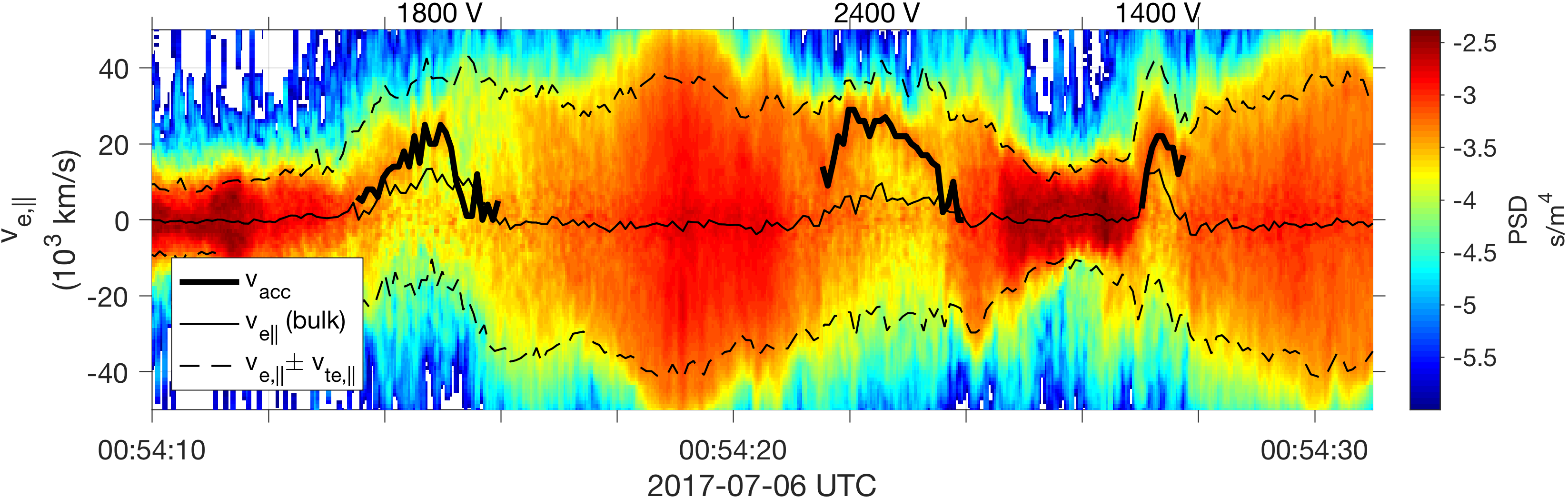}				
		\caption{Acceleration of electrons through a potential drop as seen from the reduced electron distribution. The intervals shows three acceleration channels, where the electron populations are succesively shifted towards higher energies. The thicker line follows a local maximum of the phase space density. The energy corresponding to the maximum velocity for each channel is taken as the acceleration potential $\psi$ and is written above the respective intervals. The solid thin line is the parallel electron bulk speed $v_{e,\parallel}$. The two bounding thin dashed lines marks $v_{e,\parallel}\pm v_{te,\parallel}$.}
		\label{fig:acceleration_potential_example}
	\end{figure}

	To obtain a quantitative estimate of the acceleration potential that the electrons have passed through, we investigate the reduced electron distribution in more detail.	Figure \ref{fig:acceleration_potential_example} again shows the reduced electron distribution, now for a slightly shorter time interval. The thinner black line shows the parallel electron bulk speed $v_{e,||}^{bulk}$, and the thin dashed line shows $v_{e||}^{bulk}\pm v_{te,||}$, where $v_{te,||}=\sqrt{2k_BT_{e,||}/m_e}$ is the electron thermal speed based on the parallel temperature. 
	
	To estimate the acceleration potential we use the Liouville approach and assume that a lobe population initially at rest has been accelerated to the energy corresponding to the energy where the maximum phase space density is observed, i.e. $e\psi=m_ev_{acc}^2/2$, where $v_{acc}=v_{e,||}(f_e=f_{e,max}^{local})$. To avoid picking up small variations in $f_e$ we have first applied a running average over three full 3-D distributions (the averaged distribution in Figure \ref{fig:acceleration_potential_example} can be compared to the original distribution in Figures \ref{fig:separatrix_environment}f or \ref{fig:waves_trapping_velocity}a), and thereafter again averaged $v_{acc}$ and the corresponding $\psi$ over three time steps. The obtained speeds $v_{acc}$ are shown as thick solid black lines in Figure \ref{fig:acceleration_potential_example}. We note that the phase space density between the lobe and the acceleration channels is not completely conserved, $f_e^{lb}>f_e^{acc}$. This indicates that non-adiabatic processes are at work, for example wave-particle interaction. Although beam thermalization through wave-particle interaction will decrease the average drift velocity of the beam population, it can initially tend to shift the peak phase space density to higher energies \citep[see e.g. Figure 2 in][]{Che2009}. When we find $v_{acc}$, we therefore exclude instances where the beam is thermalized beyond a certain threshold, even if it is possible to find a $f_{e,max}^{local}$. This is for example the case during the last part of the first acceleration channel. We will discuss the implications of this later. 
	
	The accelerated populations have larger speeds than the moments calculated from the entire distribution, $v^{acc}>v_{e,\parallel}^{bulk}$. This is due to the presence of an additional electron population close to $v_{e,\parallel}=0$ inside the acceleration channels. Inside the acceleration channels, the temperature (as indicated by the distance between the two dashed lines marking $v_{e,\parallel}\pm v_{te,\parallel}$) is increased relative to the lobes. This initial jump in temperature is also largely due to the presence of the additional population at $v_{e,\parallel}=0$. The population with low parallel speed could be due to wave-particle interactions, or leakage from the plasma sheet or even the lobes.
	
	The maximum potentials associated with $v_{acc}$ for the three acceleration channels are  $\psi = [1800, \ 2400, \ 1400]$ V, respectively (also written on top of Figure \ref{fig:acceleration_potential_example}). In comparison, characteristic temperatures in the lobe and plasma sheet are $T_e^{lb}=220$ eV and $T_e^{sh}=3700$ eV (obtained from the blue and red intervals shown in Figure \ref{fig:separatrix_environment}). In terms of these characteristic energies, $e\psi\approx[8, \ 11, \ 6] \ T_e^{lb}\approx[0.5 \ 0.6, \ 0.4] \ T_e^{sh}$. Note that the thermal energy of the plasma sheet is usually larger closer to the neutral sheet \citep{Baumjohann1989}. Because $|\mathbf{B}|>15$ nT during the entire shown interval, the spacecraft stay relatively close to the plasma sheet boundary layer, and therefore $T_e^{sh}$ should be considered as a lower bound.

\begin{figure}
		% +paper_electron_acceleration/phi_acc_statistics.m (new)
		% +table/separatrix_acceleration.m
		% sep/compile_acc_pot_data.m		
		\includegraphics[width=0.9\textwidth]{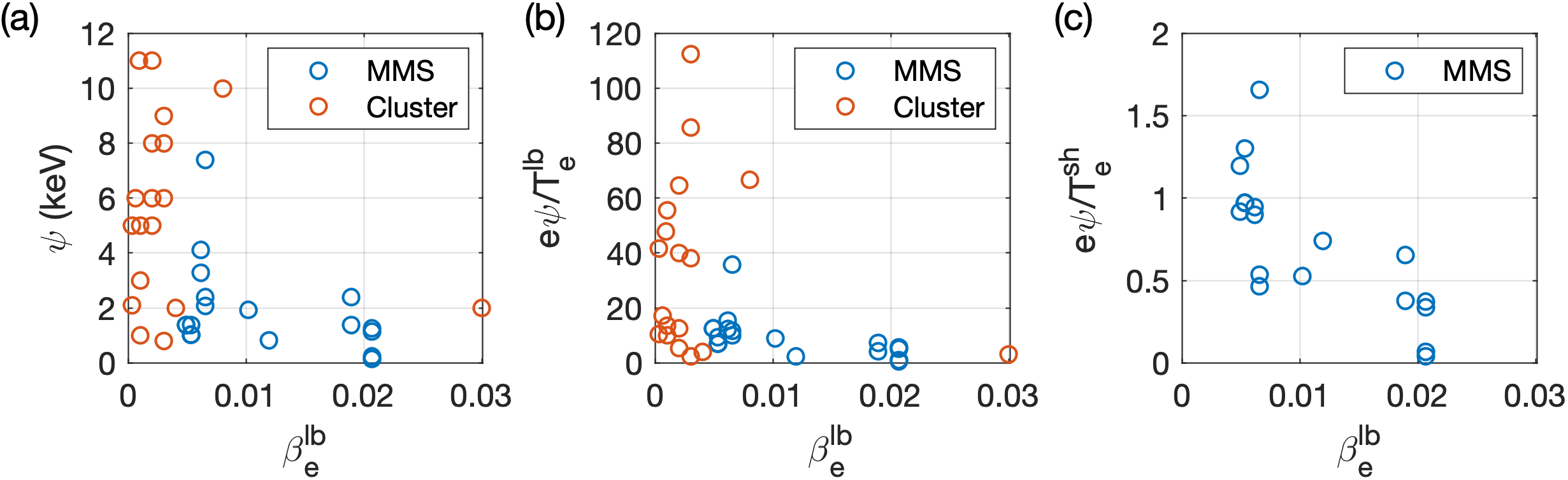}
		\caption{Summary of acceleration potential for a few electron acceleration channels, compared to previous events observed by Cluster \citep{Borg2012,Egedal2015}. Cluster results are adapted from Table 1 in \cite{Egedal2015}. (a) $\psi$ is inversely proportional to $\beta_e^{lb}$, and (b) many times the electron thermal energies per charge in the lobe $T_{e}^{lb}/e$. (c) $e\psi$ is comparable to plasma sheet thermal energies $T_e^{sh}$. Note that in events where the spacecraft only stay at the edge of the boundary layer, $T_e^{sh}$ is likely underestimated, see Table \ref{table:acc_pot}.}
		\label{fig:acceleration_potential}
	\end{figure}

	We have performed the same analysis as described above for a few other acceleration channels from a total of five burst intervals during two days in July, 2017, listed in Table \ref{table:acc_pot}. The results are shown in Figure \ref{fig:acceleration_potential} (blue circles). As before, all events we have included from MMS show clear features of a cold population being accelerated through a potential drop. We have not included flat-top distributions, or events in which the entire electron beam has likely been thermalized already. However, many events show evidence of some degree of thermalization. We show the results as a function of electron beta in the lobe $\beta_e^{lb}$. Note that the acceleration channels that are from the same burst interval can correspond to the same lobe and/or plasma sheet intervals. It is also possible that acceleration channels that are observed in close proximity to each other can be channels that are crossed multiple times. For these MMS events we find that $\psi= 1 - 8$ keV, $e\psi/T_e^{lb} = 1-15$ (with one value at 35), and $e\psi/T_e^{sh} =  0.1 - 1.7$. However for the last seven acceleration channels in Table \ref{table:acc_pot}, $T_e^{sh}$ is likely underestimated, as MMS only skirted the plasma sheet boundary layer. Regardless of this, electrons passing these acceleration channels have already reached a substantial fraction of their final energy before entering the magnetic reconnection exhaust proper. In agreement with previous results obtained by Cluster \citep{Borg2012,Egedal2015} (red circles), the acceleration potentials show an inverse dependence on $\beta_e^{lb}$. We note that these two studies seem to cover different ranges of $\beta_e^{lb}$. This could be due to selection bias, or instrumental differences related to the accuracy to which the densities and temperatures can be determined.	Another statistical study of electron distributions in magnetic reconnection regions by Cluster found electrons beams directed towards the X line with an occurrence frequency of about 20\% in the off-equatorial region \citep{Asano2008}. The beams had energies of 4-10 keV. 
	
	of their events the spacecraft observed electron beams propagating inward towards the X line. They found that the beams had a higher occurrence frequency 

	\begin{table}
		\begin{tabular}{  l  c  c  c  c } 
			\hline
			Acceleration channel time interval & $\psi$ (V) & $T_e^{lb}$ (eV)& $T_e^{sh}$ (eV) & $\beta_e^{lb}$ \\ \hline
			2017-07-03T21:54:31.700 - 54:32.700 & 3300 & 270 & 3500 & 0.006 \\ \hline
 			2017-07-03T21:54:37.200 - 54:40.600 & 4100 & 270 & 4600 & 0.006 \\ \hline
			2017-07-03T21:55:04.900 - 55:05.700 & 2100 & 210 & 4500 & 0.007 \\ \hline
			2017-07-03T21:55:06.900 - 55:08.800 & 7400 & 210 & 4500 & 0.007 \\ \hline
			2017-07-03T21:55:11.400 - 55:12.600 & 2400 & 210 & 4500 & 0.007 \\ \hline
			2017-07-06T00:54:13.900 - 54:15.600 & 1800 & 220 & 3700 & 0.010 \\ \hline
			2017-07-06T00:54:21.700 - 54:23.600 & 2400 & 330 & 3700 & 0.019 \\ \hline
			2017-07-06T00:54:26.900 - 54:27.700 & 1400 & 330 & 3700 & 0.019 \\ \hline
			2017-07-06T00:55:30.000 - 55:30.800 & 100 & 220 & 3400 & 0.021 \\ \hline
			2017-07-06T00:55:32.700 - 55:33.100 & 200 & 220 & 3400 & 0.021 \\ \hline
			2017-07-06T00:55:33.500 - 55:34.100 & 1100 & 220 & 3400 & 0.021 \\ \hline
			2017-07-06T00:55:39.900 - 55:42.100 & 1300 & 220 & 3400 & 0.021 \\ \hline
			2017-07-06T08:16:38.400 - 16:39.100 & 800 & 350 & 1100 & 0.012 \\ \hline
			2017-07-06T13:54:28.900 - 54:29.600 & 1400 & 110 & 1200 & 0.005 \\ \hline
			2017-07-06T13:54:33.600 - 54:34.300 & 1400 & 110 & 1500 & 0.005 \\ \hline
			2017-07-06T14:07:16.700 - 07:18.100 & 1000 & 150 & 1100 & 0.005 \\ \hline
			2017-07-06T14:07:18.900 - 07:19.300 & 1000 & 150 & 1100 & 0.005 \\ \hline
			2017-07-06T14:07:20.100 - 07:21.800 & 1400 & 150 & 1100 & 0.005 \\ \hline
			2017-07-06T14:07:28.200 - 07:28.800 & 1000 & 150 & 1100 & 0.005 \\ \hline
		\end{tabular}
	\caption{Acceleration potential $\psi$ obtained from a total of five burst intervals in July 2017. $T_e^{sh}$ is chosen as the largest temperature observed in the proximity of the acceleration channel.  In events where the spacecraft only stay at the edge of the boundary layer, $T_e^{sh}$ is underestimated, this is likely the case for the seven last events.}
	\label{table:acc_pot}
	\end{table}

\begin{figure}
		% +paper_electron_acceleration/phi_acc_statistics.m (new)
		% +table/separatrix_acceleration.m
		% sep/compile_acc_pot_data.m		
		\includegraphics[width=0.6\textwidth]{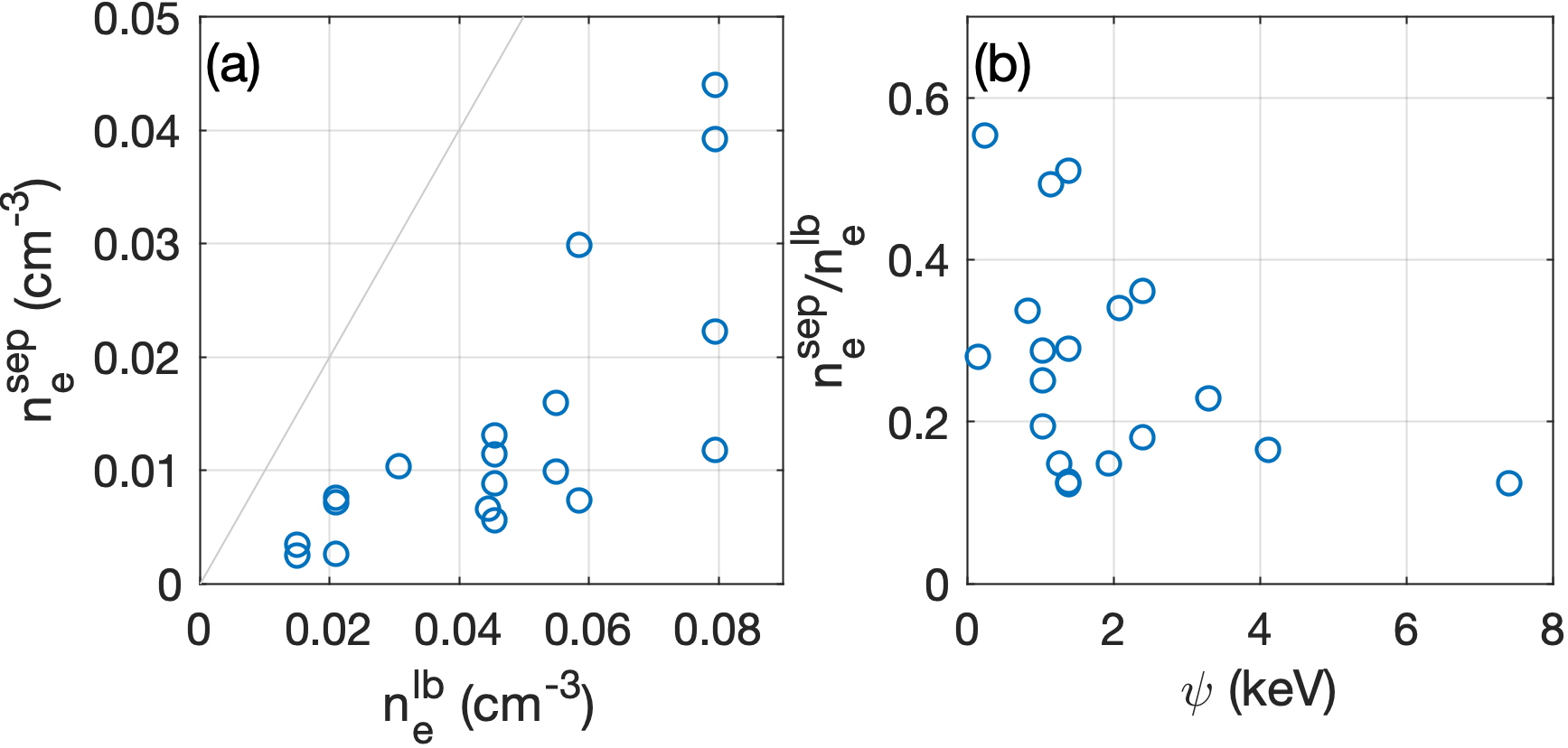}
		\caption{(a) The densities inside acceleration channels, $n_e^{sep}$, are always lower than the lobe densities $n_e^{lb}$. The gray line marks $n_e^{lb}=n_e^{sep}$. (b) The ratio of densities between the lobes and acceleration channels, $n_e^{sep}/n_e^{lb}$, show an inverse dependence on the acceleration potential $\psi$.}
		\label{fig:density_cavity}
	\end{figure}

	Similar to previous observational studies of both dayside \citep[e.g.][]{Vaivads2004b} and nightside \citep[e.g.][]{Lu2010,Wang2012} magnetic reconnection, the acceleration channels we study in this paper are associated with density cavities. Figure \ref{fig:density_cavity}a shows the relation between the lobe densities $n_e^{lb}$ and the minimum densities inside the acceleration channels $n_e^{sep}$. For all events, $n_e^{sep}<n_e^{lb}$. We do not show it here, but for all the events, the densities on the plasma sheet side of the acceleration channels are larger than both $n_e^{lb}$ and $n_e^{sep}$. Figure \ref{fig:density_cavity}b shows that the ratio of densities between the lobes and the acceleration channels, $n_e^{sep}/n_e^{lb}$, becomes smaller with increasing acceleration potential $\psi$. The decrease in density between the lobe and the acceleration channels is expected from the conservation of phase space density of an accelerated plasma population \citep[e.g.][]{Schamel1982_phys_script}. The existence of density cavities at the separatrices is also in agreement with numerical simulations of symmetric antiparallel \citep[e.g.][]{Shay2001,Lu2010,Egedal2015}, and guide-field \citep[e.g.][]{Pritchett2004} magnetic reconnection. In the case of guide-field reconnection, the reconnection electric field has a component parallel to the magnetic field. This results in enhanced parallel acceleration by the reconnection electric field, and larger density cavities at two opposing of the four separatrices. In this study, we have not differentiated between strictly antiparallel and guide-field reconnection. However, since all the events are from the tail, it is likely that any guide field, if present, is low or moderate. 
	
\subsection{Width of acceleration region}
\begin{figure}
		% paper_electron_acceleration.estimate_channel_thickness
 		\includegraphics[width=0.6\textwidth]{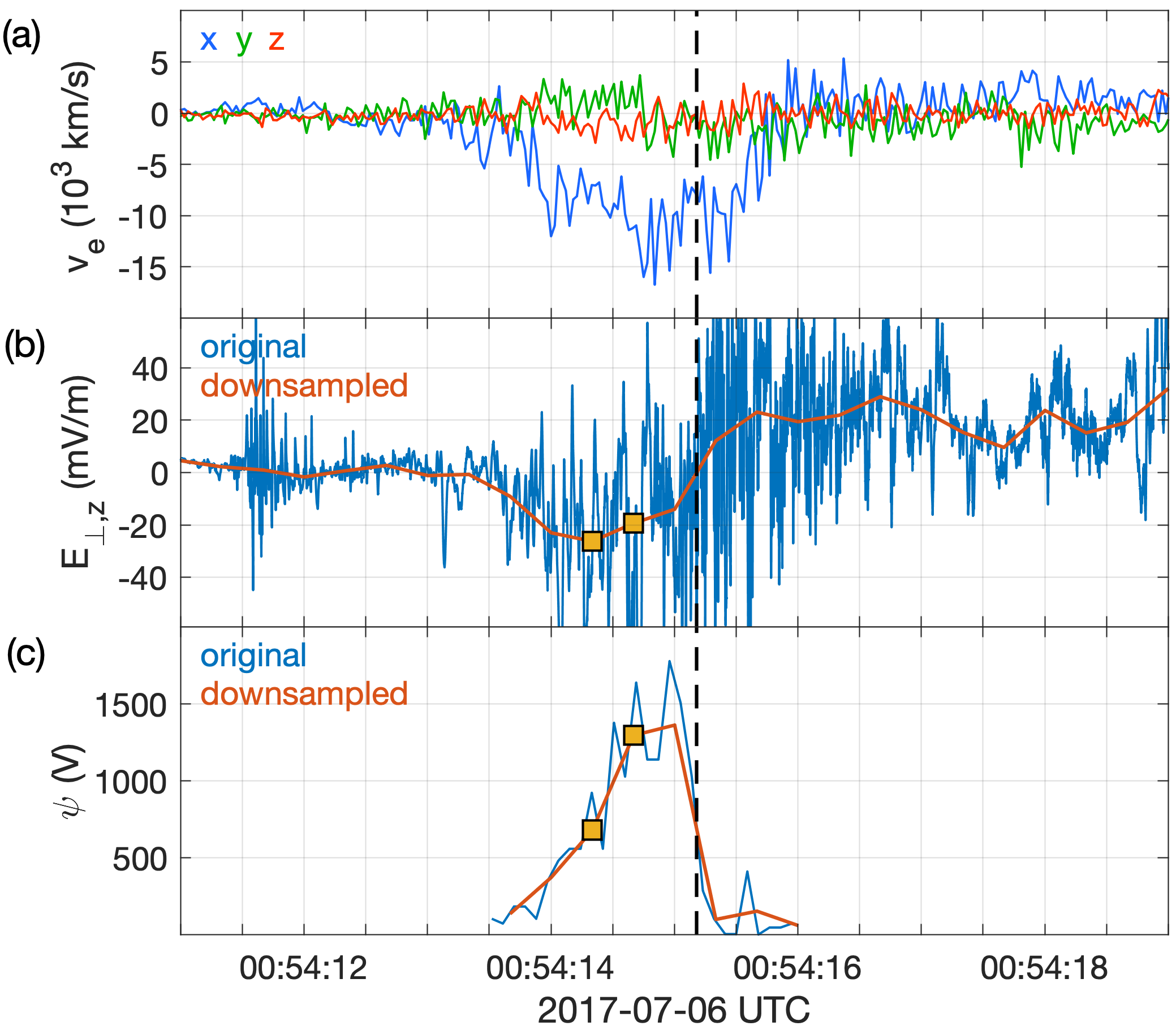}		
		\caption{Estimation of acceleration channel thickness using data from MMS1. (a) Electron flow $\mathbf{v}_e$. (b) Perpendicular electric field $E_{\perp,z}$ at original sampling rate and downsampled to 3 Hz. $E_{\perp,z}$ reverses around the time where $|\mathbf{v}_e|$ is the largest (marked by vertical dashed line). We therefore take the spacecraft to be located at the center of the acceleration channel at this time. (c) Acceleration potential $\psi$ at original cadence and downsampled to 3 Hz. Applying Eq. \ref{eq:channel_thickness} to the two times marked by yellow squares, we estimate $l_z\sim 25-65$ km, giving an acceleration channel thickness of $50-130$ km.}
		\label{fig:acc_channel_thickness}
	\end{figure}

	Before we continue, we shall make a rough estimate of the width of the first acceleration channel in Figure \ref{fig:acceleration_potential_example}, which had $\psi=1800$ V. For this event, it is not possible to reliably determine the spacecraft trajectory relative to the boundary layer from timing analysis. We therefore take a different approach using the perpendicular electric field. We make the following assumptions: (1) The acceleration potential is electrostatic in nature, and the parallel potential drop is therefore accompanied by a perpendicular potential drop. This is consistent with the divergent electric field $E_{\perp,z}$ centered around the electron flow shown in Figures \ref{fig:acc_channel_thickness}a and \ref{fig:acc_channel_thickness}b. We show the original field and the field downsampled to 3 Hz, to highlight the DC variations. We can determine that the field is divergent because MMS cross the southern separatrix from the lobe to the plasma sheet and observes a negative-positive polarity of $E_{\perp,z}$. A divergent electric field is associated with a positive electrostatic potential, consistent with the acceleration of electrons in towards the X line. This can also be seen in numerical simulations of magnetic reconnection \citep[e.g. Figures 4b-4c in][]{Divin2012}. However, some simulations also show that the electron acceleration along separatrices are in part due to inductive electric fields \citep{Egedal2015,Bessho2015}. Figure \ref{fig:acc_channel_thickness}c shows the obtained potential $\psi$ at the original cadence and downsampled to 3 Hz, like the electric field. (2) The perpendicular profile of the acceleration channel is Gaussian: $\psi_\perp(z) = \psi_0\exp(-z^2/2l_z^2)$, where $z\sim z_{GSE}$ is the coordinate perpendicular to both $\mathbf{B}$ and the main electron flow $\mathbf{v}_e$, and $\psi_0=\psi(z=0)$ is the potential in the center of the acceleration channel. While the Gaussian shape is somewhat arbitrary, we have no way to better determine the exact shape. The perpendicular electric field associated with this potential structure has peaks values $|E_{z}^{max}| = l_z^{-1}\psi_0\exp(-1/2)$ at $z=\pm l_z$, where the potential is $\psi(z=\pm l_z) = \psi_0\exp(-1/2)$. The half width is thus given by
	\begin{eqnarray} 
	l_z=\psi(z=\pm l_z)/|E_{z}^{max}|.
	\label{eq:channel_thickness}
	\end{eqnarray} 
	We now choose two time steps from where the electric field is the strongest, marked by yellow squares in Figures \ref{fig:acc_channel_thickness}b and \ref{fig:acc_channel_thickness}c. At these times, because the observed electric field maximizes here, the spacecraft are presumably located at an intermediate distance from the center of the acceleration channel, close to $l_z$. For the two points $|E_{z}^{max}| = [25,19]$ mV/m, and 	$\psi(z=\pm l_z)= [700,1300]$ V, giving estimated half widths $l_z=[25,65]$ km, respectively. The estimated thickness of the acceleration channel is thus $L=2l_z = 50 - 130$ km. In comparison, the ion and electron thermal gyroradii ranges between $100-400$ km, and $2-8$ km, respectively, where the smaller (larger) values are taken at the lobe (sheet) side of the acceleration channel. The Debye length is $\sim 0.5$ km in the lobes and reaches 4 km inside the acceleration channel. 
	
	\section{Wave activity}
	\label{sec:waves}
	Inside, and in the vicinity of the acceleration channels, large amplitude parallel electric fields are typically observed. An example is shown in Figure \ref{fig:separatrix_environment}h, where the largest amplitude fields form bipolar pulses, often termed electrostatic solitary waves (ESW) (see also Figure 5 in \cite{Huang2019}). 
	
	To quantify to what extent the electric field can affect the electrons and modify their velocity, it is helpful to look at the sum of the kinetic and potential electron energy in the frame of the wave traveling at speed $v_{ph}$: 
	\begin{eqnarray}
	U = \frac{m_e}{2}\left(v-v_{ph}\right)^2-e\phi, 
	\label{eq:U}
	\end{eqnarray}
	which is a constant of motion. If $U<0$, the electron is following a trapped trajectory, and if $U>0$, the electron is following a passing trajectory \citep{Bernstein1957}. The electrons can transition from passing to trapped trajectories (or vice versa) if the wave field is growing (or decaying) -- the electrons become trapped (or released). The limiting velocity that separates trapped and passing electron trajectories at the point where the potential is the largest,  $v(U=0,\phi=\phi_{max})$,  is
	\begin{eqnarray}
	v_{tr} = v_{ph}\pm\sqrt{2e\phi_{max}/m_e},
	\label{eq:vtrap}
	\end{eqnarray}
	called the trapping velocity. The trapping velocity is a good indication of what part of the electron distribution is likely to interact efficiently with the wave. 
	
	To find the trapping velocities, we need to find the propagation velocities $v_{ph}$, and electrostatic potentials $\phi$ of the waves. When the same wave structure is observed by two or more of the spacecraft, we can perform interferometry measurements to obtain the propagating velocity. That is, we measure the delay between the times the structure is observed by the different spacecraft, and compare it to the spacecraft separations. This is possible because the spacecraft separation of about 10 km is comparable to the typical length scale of the wave forms $\sim10 \lambda_{De}$, where $\lambda_{De}=(\epsilon_0k_BT_e/ne^2)^{1/2}$ is the Debye length \citep{Graham2016a}. Inside the acceleration channel $\lambda_{De}=3.3$ km (using $T_e\sim2000$ eV and $n_e=0.01$ cm$^{-3}$). 
	
	\begin{figure} % mms_20170706_005403.figures_selfcontained
		\includegraphics[width=0.99\textwidth]{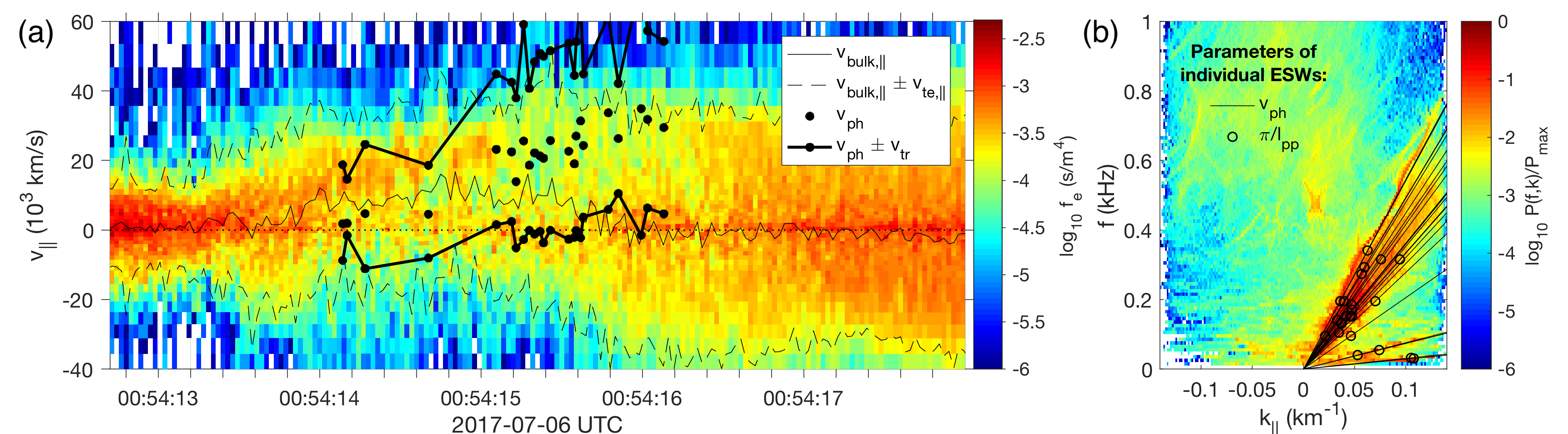}
		\caption{Interaction range of wave-field and electron population. (a) Reduced electron distribution. The thicker lines show the trapping range centered around the phase velocity (black dots). (b) Normalized power of $E_{\parallel}$ as a function of $k_\parallel$ and $f$. The lines show the phase velocities of the individual wave structures marked in panel (a). The circles show an estimate of the ESW wavenumber corresponding to a wavelength $\lambda=2l_{pp}$, $k_\circ= \pi/l_{pp}$. }
		\label{fig:waves_trapping_velocity}
	\end{figure}

	Figure \ref{fig:waves_trapping_velocity}a again shows the reduced electron distribution, now only for the time interval corresponding to the first acceleration channel. The black dots show $v_{ph}$ of individual ESWs. We find that $v_{ph}$ is loosely proportional to the velocity of the drifting electron population. Figure \ref{fig:waves_trapping_velocity}b shows the normalized power spectrum of $E_\parallel$ as a function of frequency $f$ and parallel wavenumber $k_{\parallel}$ obtained from four spacecraft interferometry for the time interval shown in Figure \ref{fig:waves_trapping_velocity}a. The method is described for two points of measurement by \cite{Graham2016a}, but here generalized to four points. This removes the need to assume a given propagation direction. The resolvable $k_{\parallel}$'s are related to the inter spacecraft separation as $k_{\parallel,max} = \pi/\max(\Delta l_{ij,\parallel})$, where $\Delta l_{ij,\parallel}$ is the distance between the individual spacecraft pairs (denoted by indices $i$ and $j$) parallel to the ambient magnetic field. In our case $\max(\Delta l_{ij,\parallel}) = \Delta l_{14,\parallel}=25$ km, giving $|k_{\parallel}|_{max}\approx0.125$ km$^{-1}$. The lower power found at $k_\parallel\lesssim0.05$ km$^{-1}$ and $f\gtrsim0.4$ kHz might be due to spatial aliasing. The black lines mark the phase velocity of the individual ESWs (as shown in Figure \ref{fig:waves_trapping_velocity}a). While the general trend is that the the phase velocities increase towards the plasma sheet, it is possible to roughly divide the ESWs into two groups based on their speeds: one earlier group with lower speeds and lower $f$ and one later group with higher speeds and higher $f$. 
	
\begin{figure}
		\includegraphics[width=0.45\textwidth]{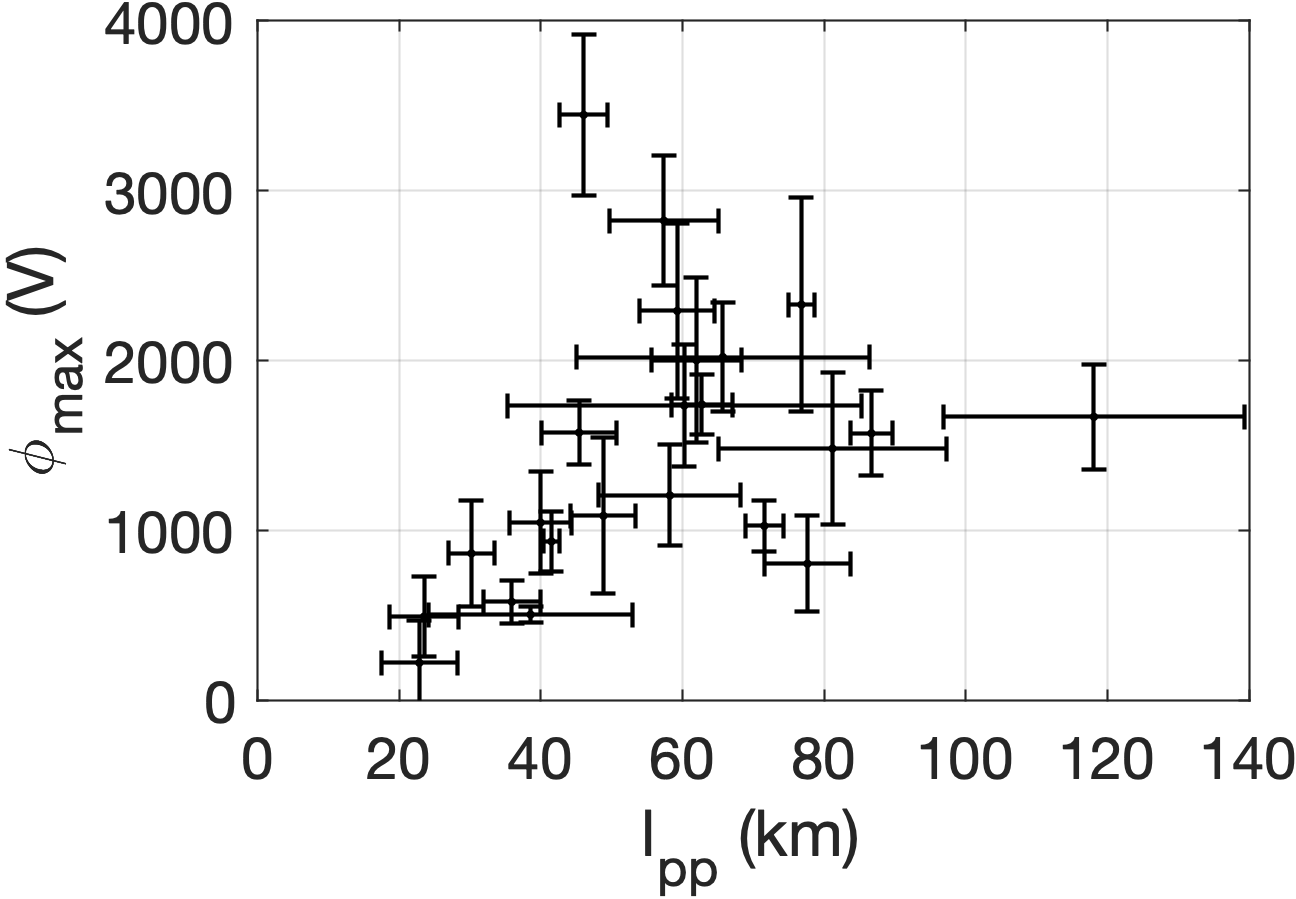}
		\caption{Electrostatic potential and peak-to-peak length scales of ESWs. Since each spacecraft can pass along a different trajectory through the ESW, the four spacecraft can observe different $l_{pp}$ and $\phi_{max}$, respectively. We therefore show the standard deviation centered on the mean value for each ESW. The average values for the entire group of ESWs are $\left< l_{pp}\right>=57$ km, and $\left<\phi_{max}\right>=1500$ V.}
		\label{fig:phi_lpp}
	\end{figure}

	Using $v_{ph}$, we can obtain the distances between the positive and negative peaks of the bipolar electric fields, which we call the peak-to-peak length scale. We find that they vary between $l_{pp}=20$ km and $l_{pp}=120$ km with a mean value of $\left< l_{pp}\right>=57$ km (Figure \ref{fig:phi_lpp}). The wavenumbers corresponding to wavelengths $\lambda=2l_{pp}$ of the individual ESWs, $k_{\circ} \sim \pi/l_{pp}$, are marked by circles in Figure \ref{fig:waves_trapping_velocity}. For some of the ESWs, the electric field perpendicular to the ambient magnetic field was comparable or even larger than the parallel field (not shown). This suggest the perpendicular length scales can be comparable to the parallel length scales \citep{Franz2000}. A large range of estimated $l_{pp}$'s are comparable to the estimated width of the acceleration channel, which was $L \approx 50-130$ km. 
	
	The potential of the waves along the trajectory of the spacecraft are calculated by integrating the parallel electric field, using the parallel component of the measured phase velocity, ${dl = -v_{ph}dt}$:	
	\begin{eqnarray*}
	\phi = \int E_{\parallel}v_{ph}dt.
	\end{eqnarray*}
	The maximum electrostatic potential of each ESW along the spacecraft trajectory $\phi_{max}$ varies between $\phi_{max}=300$ V and $\phi_{max}=4000$ V with a mean value of $\left< \phi_{max}\right>=1500$ V (Figure \ref{fig:phi_lpp}). The corresponding trapping speeds $v_{tr}$ (Eq. \ref{eq:vtrap}) are shown in Figure \ref{fig:waves_trapping_velocity}a as two black lines bounding $v_{ph}$. At earlier times where $v_{ph}<5\times10^3$ km/s, the trapping range encompasses significant parts of the beam, indicating favorable conditions for strong wave-electron interaction. At later times where $10\times 10^3<v_{ph}<35\times10^3$ km/s, the beam is not as apparent and has likely become significantly thermalized, also indicative of strong wave-particle interaction. 
	
	Note that at the later part of the interval, where $v_{ph}$ are larger, the beam has become significantly thermalized. Since we did not take into account significantly thermalized beams when determining $\psi$, the maximum acceleration speed obtained for this acceleration channel ($v_{acc}\sim25\times10^3$ km/s, corresponding to $\psi=1800$ V), is smaller than the observed phase velocities, $v_{acc}<v_{ph}$.
	
\begin{figure} % mms_20170706_005403.figures_selfcontained
		\includegraphics[width=0.95\textwidth]{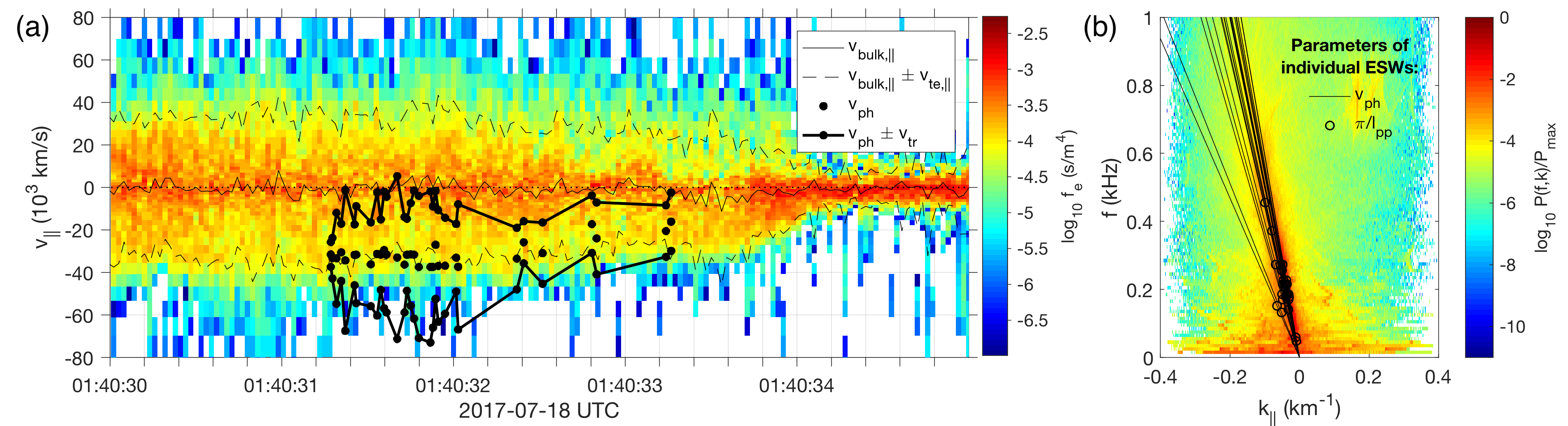}
		\includegraphics[width=0.95\textwidth]{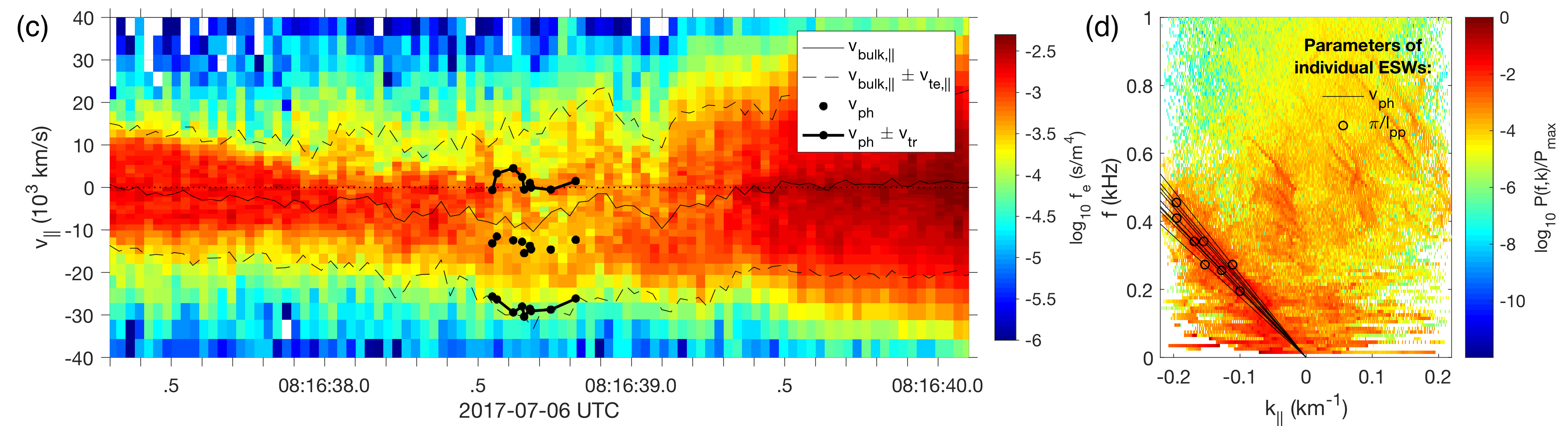}
		\caption{Interaction range of wave field and electron population. (a), (c) Reduced electron distribution. The thicker lines show the trapping range centered around the phase velocity (*). (b), (d) Normalized power of $E_{\parallel}$, as a function of $k_{||}$ and $f$. The lines show the phase velocities the individual wave structures marked in panel (a). The circles show an estimate of the central wave number $k_o\sim \pi/l_{pp}$ assuming the wavelength is given roughly by $\lambda=2l_{pp}$. While only the second case shows a distinct beam, both cases show $v_{ph}$'s that are proportional to the electron energy and interaction ranges $v_{ph}\pm v_{tr}$ that covers a significant part for the electron distributions.}
		\label{fig:waves_trapping_velocity_more_examples}
	\end{figure}

	We have performed this analysis for a number of events, and present two more of them in Figure \ref{fig:waves_trapping_velocity_more_examples}. We observe both similarities and differences between the different cases. For example, in Figure \ref{fig:waves_trapping_velocity_more_examples}a, although there is an asymmetry in $f_e(v_\parallel)$ close to the lobe towards the end of the interval, the electron bulk velocity is close to zero, and no distinct beam is observed. However, the phase velocities are proportional to the energy at which the phase space density begins to decrease rapidly (we refer to this energy as the shoulder energy). This might indicate that the beam has already been destroyed by the wave-electron interaction, and that what we observe is the thermalized beam. This is supported by the fact that the trapping range $v_{ph}\pm v_{tr}$ covers a large part of the electron distribution. Because no distinct beam is observed, this time interval is not included in Table \ref{table:acc_pot} or Figure \ref{fig:acceleration_potential}. Figure \ref{fig:waves_trapping_velocity_more_examples}c shows an acceleration channel with a distinct beam, which is included in Table \ref{table:acc_pot} and Figure \ref{fig:acceleration_potential}. Again we find that the phase velocities are proportional to the beam speed and that the wave interaction range covers the beam. Figures \ref{fig:waves_trapping_velocity_more_examples}b and \ref{fig:waves_trapping_velocity_more_examples}d show that the phase velocities obtained from timing analysis of individual ESWs correspond well to the maximum wave power in the dispersion relation obtained form four spacecraft spectral analysis. In Figure \ref{fig:waves_trapping_velocity_more_examples}d, we can see clear indications of spatial aliasing.	
	
	\section{Spatiotemporal evolution and instability analysis}
	\label{sec:instability}
	To investigate whether the observed plasma distributions can account for the generation of the ESWs, we solve the unmagnetized, electrostatic dispersion equation:	
	\begin{eqnarray}
		0 = 1-\sum_s \frac{\omega_{ps}^2}{k^2v_{ts}^2}Z'\left(\frac{\omega-kv_{ds}}{kv_{ts}}\right)
		\label{eq:dispeq}
	\end{eqnarray}		
	for the event presented in Figure \ref{fig:waves_trapping_velocity}. Here, $Z$ is the plasma dispersion function, and '$s$' denotes the different plasma populations. Based on the observed ESW characteristics, the ESWs in Figure \ref{fig:waves_trapping_velocity} can be divided roughly into two groups. One group with lower $v_{ph}$ observed at earlier times and one group with larger $v_{ph}$ observed at later times. We will therefore investigate different combinations of plasma populations. 
	
\begin{figure}
		\includegraphics[width=0.98\textwidth]{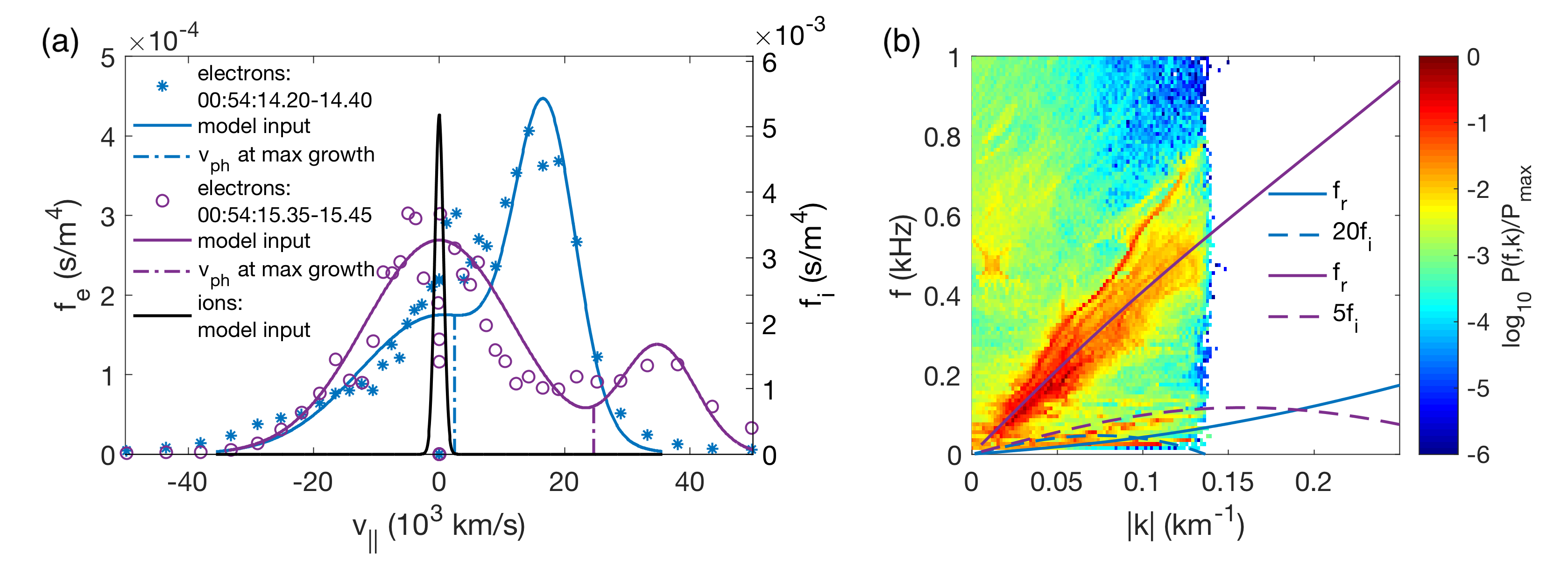}
		\caption{Wave instability analysis. (a) Observed and fitted model distribution during two time intervals corresponding to where the slower (blue) and faster (purple) ESWs are observed, respectively. (b) Observed dispersion relation and real ($f_r$) and imaginary ($f_i$) frequencies obtained by solving Eq. \ref{eq:dispeq} for the two distributions in (a). Both solutions show real frequencies corresponding well to the observed ones. For the slow ESWs the range of $k_{\parallel}$ with positive growth rates $f_r>0$ corresponds to where the wave power is the largest. For the faster ESWs, the peak growth rate is shifted towards larger $k_\parallel$, or equivalently smaller wavelengths.}
		\label{fig:disp_rel}
	\end{figure}

	We expect the accelerated lobe electron population to be the main driver of the instability. This beam, however, can be in different stages of evolution. Further downstream of the acceleration channel, observations and simulations show that the peak phase space density is shifted toward larger speeds, but also that the beam is weaker, i.e. it has a lower density relative to the electron population at lower speeds (compare the blue stars and purple circles in Figure \ref{fig:disp_rel}a). To the zeroth order, the shift to larger speed is due to the acceleration. However, the wave-particle interaction can also contribute to this as follows: electron trapping removes phase space density from the lower speed edge of the beam, decreasing its density and at the same time shifting the beam peak to higher speeds \cite[e.g.][]{Che2009}. Recall that when acquiring the acceleration potential $\psi$ in section \ref{sec:acceleration_potential}, we did not include significantly thermalized beams. The distribution shown with purple circles in Figure \ref{fig:disp_rel} is an example of such a distribution that was considered too thermalized. Note that a beam drift speed of $35000$ km/s would correspond to an acceleration potential of 3500 eV. The electron population at lower speeds can be plasma sheet electrons that enter the acceleration channel during part of their gyromotion. It can also be the trapped electrons that were originally part of the beam. A large part of the low-velocity electrons are within the trapping range of the ESWs. The ions can be both cold lobe and hotter plasma sheet populations. However, the ion thermal speeds of both lobe and plasma sheet ions are both low in comparison to electron and phase speeds, so in the dispersion analysis we will only consider a single (medium hot) ion population with a temperature of $T_i = 5$ keV, corresponding to a thermal speed of $v_{ti}=980$ km/s. An ion temperature of 10 keV would correspond to a thermal speed of 1380 km/s, which is not a significant difference considering the phase speeds and electron speeds. 

	We consider two different electron distributions, based on observed distributions at slightly earlier (blue) and later (purple) times where the observed $v_{ph}$ are slower and faster, respectively (Figure \ref{fig:disp_rel}a). The solid lines show fits to the 1-D reduced electron distributions (blue: $n_{e} = [0.055,0.045]$ cm$^{-3}$, $T_e = [900,130]$ eV, $v_d=[0,17000]$ km/s, purple: $n_{e} = [0.080,0.020]$ cm$^{-3}$, $T_e = [800,200]$ eV, $v_d=[0,35000]$ km/s) and are used as input to Eq. \ref{eq:dispeq}. The resulting real and imaginary frequencies $f = f_r + if_i$ obtained from Eq. \ref{eq:dispeq} are shown overlaid with the observed power distribution in Figure \ref{fig:disp_rel}b. For both cases we obtain good matches to the real frequencies $f_r$, while the maximum growth for the faster ESWs are shifted towards higher $k_\parallel$'s. The growth rate for the faster beam is about ten times larger than that for the slower beam. The phase speeds at maximum growth rate are $v_{ph}^{slow} = 2500$ km/s and $v_{ph}^{fast} = 27000$ km/s, respectively, shown as vertical dashed-dotted lines in Figure \ref{fig:disp_rel}a.
	
	For the slower ESWs, depending on small variations of the input parameters, either the ion-electron or electron-electron modes are dominating. The ion-electron mode is essentially a Buneman type mode with a hot electron background that does not interact with the drifting electron population \citep{Norgren2015b}. The slow electron-electron mode is an electron-acoustic wave. The faster ESWs are generated by an electron beam-mode instability with close to constant phase speed regardless of wavenumber. The evolution of instabilities from Buneman to electron beam-mode is similar to what was described by \cite{Che2009} for guide field reconnection. Although the electron beams we study here are not located inside the EDR, the local dynamics can be the same.
			
 	Since we observe large amplitude highly nonlinear localized structures, we are not observing the waves in the linear stage of instability. The predicted growth is therefore not necessarily expected to coincide with the range of $k_{\parallel}$'s where the linear growth rate maximizes. Although some simulations do show good correspondence between observed wave characteristics and linear instability growth rates \citep[e.g.][]{Fujimoto2006,Chen2015_sepwaves}, ESWs are for example known to merge with each other and grow in size \citep[e.g.][]{Mottez1997}, which would correspond to a shift to smaller $k_{\parallel}$'s. If we would assume that the wave power at the initial stages of the beam-mode instability peaked at $0.15$ km$^{-1}$ but due to coalescence has shifted to $0.05$ km$^{-1}$, this would correspond to a change from a wavelength $\lambda=40$ km to a corresponding length scale of the ESWs of 120 km. Another effect that may play a role in modifying the wave growth is the velocity shear and the perpendicular structure of the flow channel inside which they grow \citep{Che2011b}. The waves can not be considered as plane waves with infinite extent in the perpendicular direction, which Eq. \ref{eq:dispeq} assumes. However, extending the wave analysis to include these effects is beyond the scope of this paper. ESW's are also known to be limited in size by transverse instabilities \citep{Muschietti2000,Graham2016a}. If the bounce frequency of electrons trapped in the potential well of the ESWs exceeds the gyrofrequency, the ESWs tend to become unstable and dissipate. The bounce frequency for a given potential increases with decreasing length scale, or correspondingly larger $k$'s. We have examined this relation here (not shown) and find that the limiting $k$'s are about three times larger than the $k$'s where the wave power maximizes. We therefore do not think this is the deciding factor in determining the range of observed length scales.
	
	\section{Discussion}
	We have investigated the electron acceleration and wave-particle interaction in acceleration channels located at magnetic reconnection separatrices in the magnetotail. Generally, we found that the lobe populations were gradually accelerated up to a significant fraction of the thermal energies in the plasma sheet. Nonlinear ESWs that were observed at the same time as the accelerated populations had large enough potentials to interact with a significant part of the electron distribution, including the beam. Here we will discuss how the wave-particle interaction and spatial effects are expected to alter the beam and thereby our estimate of the acceleration potential $\psi$. We will also discuss how the wave properties can be related to the thermalized beam.
	
	The beam represents the free energy of the system. When the waves have grown to amplitudes such that they are able to trap the beam, the beam can become thermalized, and the free energy source is removed. The wave trapping range at saturated wave growth is therefore expected to encompass the beam responsible for the instability growth. This is consistent with our observation: when a distinct beam could be observed, such as seen at earlier times in Figure \ref{fig:waves_trapping_velocity}a or in Figure \ref{fig:waves_trapping_velocity_more_examples}c, the wave trapping range encompassed a large part or the entirety of the beam: 
	\begin{eqnarray}
	|v_{ph}|<|v_{acc}|<|v_{ph}+v_{tr}|.
	\label{eq:deduce_vacc}
	\end{eqnarray}
	The waves observed at the same time as the beams were highly nonlinear solitary structures, indicating they could be close to saturation.
	
	When we estimated the acceleration potential $\psi$, we studied the beam component of the reduced electron distribution. However, the wave-particle interaction can affect the beam and thereby affect our estimate of $\psi$. For example, electrons will locally be accelerated to higher energies in the potential $\phi$ of the wave. Since the sampling rate of FPI is lower than the time scale of an individual ESW, the beam will appear spread out in velocity space, and the peak of the spread out beam energy will appear at higher energies than what was accounted for by the accelerating potential $\psi$. The waves can also eventually trap electrons, and thus gradually remove electrons from the lower speed edge of the beam. This will further spread out the beam in velocity space and shift the peak phase space density of the beam to higher energies. However, these thermalization effects occur gradually, and can have affected the beam significantly even before the beam has reached its final energy as governed by the larger scale potential $\psi$. 
			
	For the case presented in Figure \ref{fig:waves_trapping_velocity}a, we could clearly see that the closer to the plasma sheet the accelerated population were observed, the weaker, or more thermalized, the beam became. In the region with higher velocity ESWs, it is hard to perceive the beam at all. Since we did not take into account significantly thermalized beams when determining $\psi$, the maximum acceleration speed obtained for this acceleration channel ($v_{acc}\sim25\times10^3$ km/s, corresponding to $\psi=1800$ V) was smaller than many of the observed phase velocities, $v_{acc}<v_{ph}$. However, due to the nature of the generating instabilities, where the phase speeds fall in between the two populations, it is reasonable to assume that the speed of the unthermalized beam was larger than $v_{ph}$ as shown in the inequality in Eq. \ref{eq:deduce_vacc}. For example, for the instability analysis of the faster ESWs in Section \ref{sec:instability}, the drifting component of the electron distribution had a speed of $35\times10^3$ km/s. Therefore, although the beam-wave interaction can make the beam appear at higher energies than what is accounted for by $\psi$, the observed phase speeds should give a minimum value. This is an indication that the maximum potential drop for this acceleration channel was in fact larger than what we estimated in Figure \ref{fig:acceleration_potential_example}. In Table \ref{table:acc_pot} and Figure \ref{fig:acceleration_potential} $\psi$ could therefore be considered as lower bounds.
	
	Like mentioned above, for the beam in Figure \ref{fig:waves_trapping_velocity}a, the beam thermalization became more prominent when the beam had already been accelerated to larger energies. The point in time when the beam seemed to become significantly thermalized coincided roughly with the appearance of high velocity ESWs. It is possible that the increased thermalization could be related to the generating instabilities. The electron beam-mode instability, which we found could be responsible for the wave generation in the large velocity region had a growth rate roughly ten times larger than the instabilities that were active when the beam had lower speeds. To some extent, the increased thermalization should also be due to the integrated effect of wave-particle interactions along the acceleration channel. That is, the further down the acceleration channel the beam has progressed, the longer distance the wave-particle interaction has had the time to affect the beam. 
	
	In other cases where no distinct beam could be observed (we presented one such case in Figure \ref{fig:waves_trapping_velocity_more_examples}a), the phase speeds were instead proportional to the energy at which the phase space density started to decrease rapidly. Due to the similar wave behaviour between the cases with and without distinct beams, it is likely that the presumed beam in the latter case had already been completely thermalized. In such cases where no beam can be identified, but the wave characteristics can be obtained, it could be possible to use the inequality in Eq. \ref{eq:deduce_vacc} to indirectly estimate the beam characteristics and therefore the amount of electron acceleration as well as thermalization. Inferring the characteristics of the generating electron beam after it has already been thermalized has been done in previous studies \cite[e.g.][]{Norgren2015b}. Here, however, we have here tracked the continuous change in the beam, accompanied by the continuous change in wave characteristics, certifying the credibility of such an approach.

	We also consider the case where the weakened beam could be partially due to spatial effects perpendicular to the magnetic field. For example, if we assume that the initial lobe population is isotropic, electrons with pitchangles close to $90^\circ$ would cover a larger perpendicular distance throughout their gyromotion than electrons with pitchangles closer to $0^\circ$ or $180^\circ$. The beam would therefore be weaker at the edges of the acceleration channel than at the center. However, the gyroradii of lobe electrons, based on $T_{e}^{lb}=220$ eV and $B^{lb}=20$ nT is $\rho_e=2.5$ km, which is significantly smaller than the estimated thickness of the acceleration channel $L \approx 50-130$ km. For this gyration effect to be important, it is likely that some prior heating and pitch angle scattering would have to had taken place. 	
	
	\section{Conclusions}
	In this study we investigated the electron acceleration and subsequent wave generation and wave-particle interaction at magnetic reconnection separatrices in the magnetotail. We summarize our conclusions below.%Below, summarize our conclusions based on a total of 20 acceleration channels, and thereafter based on three events presented in more detail.
	
	\begin{itemize}
		\item Adjacent to the reconnection exhaust, we found relatively thin regions of electron lobe populations accelerated towards the X line. The electrons were accelerated to energies of 1-7 keV, several times the thermal energies within the lobe, and a significant fraction of the thermal energies inside the outflow.
		\item All acceleration regions were associated with density cavities. The decrease in densities from the lobes to the acceleration channels increased with increasing acceleration potential, consistent with theoretical predictions.
		\item For two acceleration channels presented in more detail, we could observe how the lobe populations were gradually accelerated. For one of them, the resulting beam became significantly weaker closer to the plasma sheet. 
		\item Electrostatic solitary waves observed in the acceleration regions had phase speeds proportional to the beam speeds. The potentials of the waves were large enough such that the waves could interact efficiently with a large part of the electron population, including the beam. This indicates that the waves play an important role in controlling the evolution of the beam, aiding to thermalize it.
		\item For one acceleration channel we investigated the instability of the evolving electron distribution and found that it could account for the observed wave properties. When the beam had been accelerated to moderate speeds, it was unstable to a combination of competing  Buneman and electron acoustic instabilities, generating waves at low phase speeds. When the beam had been accelerated to larger speeds and had become weaker, the distribution was unstable to an electron beam-mode instability, generating waves at larger speeds. 
	\end{itemize}
	For the events that we have examined, these wave-particle interaction ranges presented above represent the typical conditions when ESWs are observed. This shows the efficiency with which waves are able to turn directed drift energy into thermal energy of the plasma. These results are not only applicable to magnetic reconnection, but to any process in which superthermal electron beams form.

\end{document}